\documentclass[9pt,twocolumn]{article} 
\usepackage{simpleConference}
\usepackage{times}
\usepackage{graphicx}
\usepackage{amssymb}
\usepackage{url,hyperref}
\usepackage{amsmath}

\begin{document}

\title{Low power (mW) nonlinearities  of polarization maintaining fibers}

\author{Hanieh Afkhamiardakani, Luke Horstman, Ladan Arissian, and Jean-Claude Diels \\
\\
\small Center for High Technology Materials, University of New Mexico, Albuquerque, NM 87106, USA \\
\small \today
\\
\em \small corresponding author: jcdiels@unm.edu  \\
}

\maketitle
\thispagestyle{empty}

\begin{abstract}
{Polarization maintaining (PM) fibers are meant to maintain linear polarization along a preferred axis.
A PM fiber can be seen as the fiber version of a very high order waveplate, designed with different refractive indices along two orthogonal axes.
It is shown that monitoring the polarization of  initially circularly polarized light sent
through a PM fiber, leads not only to
new sensing methods, but also to power control, saturable absorption, and optical path stabilization.
  Even at peak power levels not exceeding a few mW,
nonlinear transmission is detected, with time constants in  the microsecond range.  }
\end{abstract}

\section{Introduction}

Daniel Colladon initiated the field of optical waveguiding by demonstrating the possibility of guiding light through a curved stream of water in 1841.
By 1953 image transmission through the first fiberscope was demonstrated~\cite{Hopkins54}.
Today, optical fiber has become an integral part of many fields including telecommunication~\cite{Sharma13},
medical~\cite{Correia18}, and metrology~\cite{Afkhamiardakani19b}.

Fiber lasers and fiber sensors have attracted attention in recent years due to their abilities
to provide same quality features, such as  power, linewidth, repetition
 rate, pulse duration, and sensitivity as their free-space counterparts
  while being cheaper, more compact, and easier to use.
 Passive fiber sensors are usually implemented as Sagnac interferometer ~\cite{Lv15,Shao16}, Michelson interferometers~\cite{Li14}, Fabry-Perot interferometers~\cite{Yu18}, or microfibers~\cite{WangPengfei15}
 to measure magnetic fields, strain, torsion, and temperature.
 These sensors monitor their observable of choice by monitoring the phase shift or spectral changes  of transmitted broadband light, which requires a complicated setup or specialty fiber~\cite{Dong07,Yu18}.

It is shown here that simply monitoring the polarization of the transmitted light through a polarization maintaining (PM) fiber
leads not only to
new sensing methods, but also to power control, saturable absorption, and possibility of optical path stabilization.
  Even at peak power levels not exceeding a few mW,
nonlinear transmission is detected, with time  constants in  the microsecond range.  All effects related to the Kerr nonlinear index can be
neglected in the range of powers considered here.

\section{polarization maintaining fibers}
\label{PM Fibers}
Single mode (SM) fibers exhibit some birefringence, typically stress induced, such that the polarization of a beam sent through the fiber varies with positioning and bending of the fiber.
This effect has been exploited for generating short pulses through  polarization mode-locking~\cite{Matsas92}.
Polarization maintaining fibers were introduced in order to maintain linear polarization along a preferred axis.
A PM fiber can be seen as the fiber version of a very high order waveplate: instead of the cylindrically symmetric SM fiber,
it is designed with different indices of refraction along two orthogonal axes (the ``slow axis'' for the higher index, the ``fast axis''
along the direction of lower index).  One defines the ``beat length'' (typically a few millimeters) as the distance over which the retardation
between the slow and fast light is equal to $2\pi$.  Only linear polarization launched along one of these axes is maintained.
Any other input polarization will be periodically modified along the fiber.  In PM fiber, unlike the SM fiber,
the slow and fast axes are fixed along the fiber, and the beat length is considered to be a constant.
However, small variations of beat length can take place because of environmental conditions (temperature, stress, magnetic field) or power variations of the propagating light.
By accumulating  beat length variations over long distances, we demonstrate extreme sensitivity in the measurement of
many parameters affecting the fiber birefringence.

\section{Polarization ellipse measurements}
\label{pol_ellipse}

\begin{figure}[h!]
\centering
\includegraphics[width=\linewidth]{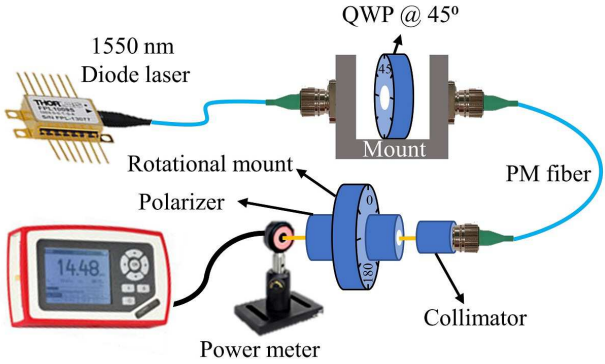}
\caption{\small Experimental setup for polarization measurement of the transmitted circularly polarized light through the PM fiber at different temperatures or powers of light. QWP: quarter wave plate}
\label{setup}
\end{figure}

\begin{figure}[h!]
\centering
\includegraphics[width=\linewidth]{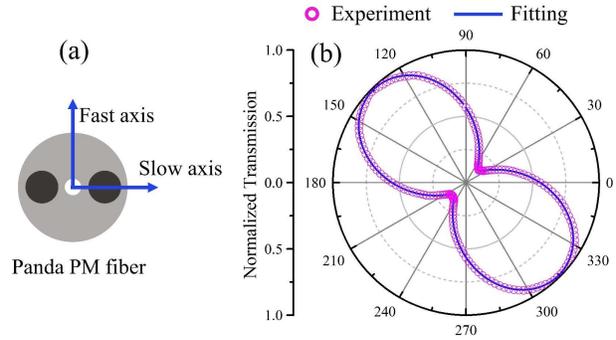}
\caption{\small a) Cross section of a PANDA PM fiber. b) An example of elliptically polarized light with  ellipticity of 0.37 and angle of 135$^\circ$ with respect to the slow axis of the PM fiber.  }
\label{expVSfit}
\end{figure}

In order to make a comprehensive determination of the polarization modification,
the complete polarization ellipse is measured for each value of a given parameter
effecting the beat length.  The setup to measure the change in polarization
after passage through the PM fiber is shown in Fig.~\ref{setup}.
A laser diode at 1550 nm generates low power, continuous wave, linearly polarized light which is sent to a quarter waveplate at 45$^\circ$
to create circularly polarized light, which excites both modes of the PM fiber independently of the
orientation of the input end of the PM fiber.  Fibers of the PANDA type were used, where the
principal axis is defined by ``stress rods'', as illustrated in Fig.~\ref{expVSfit}(a).
The transmitted light is collimated and sent through a polarizer mounted on a LABVIEW-controlled rotational mount.
The polarization measurement is performed by measuring the transmitted power ($P_t$) 
through the rotating polarizer which has a transfer function of the form of Jones matrix of
M=
$
\begin{bmatrix}
\cos^{2}\theta & \sin\theta \cos\theta \\
\sin\theta \cos\theta & \sin^{2}\theta
\end{bmatrix}
$
where $\theta$ is the angle between the transmission axis of the polarizer and the slow axis of the PM fiber in Fig.~\ref{expVSfit}(a).
 As the polarizer is rotated from 0 to 360 degrees, the power meter measures the projection of the polarization transmitted by the fiber
along the transmission axis of the polarizer~\cite{Afkhamiardakani16}.  Polarization measurements are performed
at angular increments of 2$^\circ$.  The ellipticity and the angle of the polarization ellipse are extracted by fitting
the projection data.
The accuracy of the reconstruction of the polarization state is
illustrated by the comparison between the {normalized} raw  data and the fitted ellipse projection as seen in Fig.~\ref{expVSfit}(b). {The normalized fitted projection has been used to show the following results.}
\section{Power dependent polarization changes}
\subsection{Measurements}
\label{power}

 \begin{figure}[t]
\centering
\includegraphics[width=\linewidth]{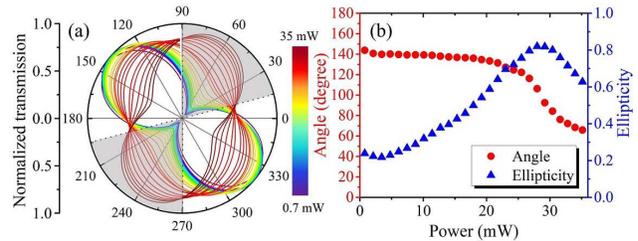}
\caption{\small a) Projection patterns of the transmitted circularly polarized laser light at different powers through a
 17.5 cm PM fiber (at room temperature) followed by a rotating polarizer. 
  b) Ellipticities and angles of the polarization ellipses associated to (a).}
\label{polVSpower}
\end{figure}

\begin{figure}[h!]
 \centering
 \includegraphics*[width=\linewidth]
{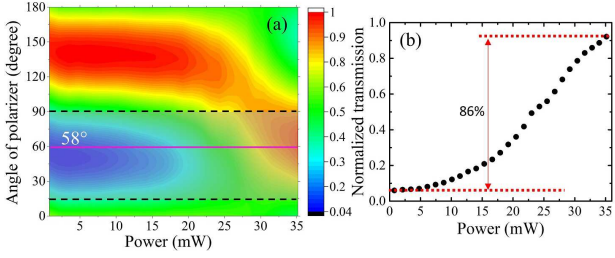}
\caption[]{\small a) Color coded transmission of circularly polarized light at
 different powers through a 17.5 cm PM fiber followed by a
  rotating polarizer. b) Transmission versus power of 
  circularly polarized light for the optimum polarizer angle of 58$^{\rm o}$ shown in (a).}
 \label{saturation}
\end{figure}

Intensity dependent polarization changes are well known in fibers.  The Kerr effect induced change of index $n_2I$
is the origin of mode-locking by polarization rotation.  Since the nonlinear index in silica is of the order of $10^{-16}$ cm$^2$/W,
this mode-locking technique involves peak powers of the order of tens of watts.  Instead, our measurements are
performed with continuous radiation in a much lower power range.
Figure~\ref{polVSpower} shows the polarization modification for a 17.5 cm PM fiber, as the
power is incremented from 0.7 to 35 mW.  The data presented in Fig.~\ref{polVSpower}(a) show a  polarization ellipse
rotating as the power of
circularly polarized light input to the PM fiber is incremented.
The change in ellipticity and angle of the ellipse are plotted as a function of input power in
Fig.~\ref{polVSpower}(b).  Even with such a short fiber section, the polarization state
is sensitive to a mW change in light power.
{Figure~\ref{saturation}(a) shows the  normalized transmission through the PM fiber followed by a polarizer at different angles,
as a function of the input  power.
At the polarizer angle of 58$^\circ$, the change in transmission with increasing power
reaches 86\% [Fig.~\ref{saturation}(b)].
Therefore a short piece of PM fiber can be used as a saturable absorber
of very low saturation power.
This nonlinearity  can be exploited for making pulsed lasers and sensors of very
small energy consumption.  It would also avoid the  nonlinearities that affect
 the group and phase velocities in fiber based sources
 of frequency combs~\cite{Afkhamiardakani19b}.
The grey regions in Figs.~\ref{polVSpower}(a) and \ref{saturation}(a) correspond to the range of polarizer angles
 (from 15$^\circ$ to 90$^\circ$) for which the saturable absorption is observed.}

 \subsection{Interpretation of the power dependence}
\label{interpretation}
\begin{figure}[t]
\centering
\includegraphics[width=.7\linewidth]{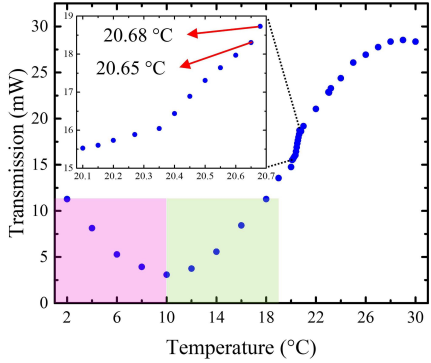}
\caption{\small Transmission of circularly polarized light at 30.4 mW sent to the PM fiber at different temperatures followed by a polarizer oriented at
46 degrees which is shown by the dashed lines located at 46$^\circ$ in Fig.~\ref{polVSpower}(a) and (c). Temperature determination is ambiguous in the highlighted regions. Inset: enlarged scale to show the sensitivity of the temperature sensor. }
\label{trans-temp}
\end{figure}

At such low power levels, power dependent index change could be
 caused by thermal effects.
  The minuscule absorption of 0.5 dB/km  in the silica fiber is sufficient
to raise the temperature of the core to modify the stress-induced birefringence of the
fiber.  It has been reported previously that the birefringence of  PM fibers
is very sensitive to temperature~\cite{Zhang93,Eickhoff81}. 
To verify that the same temperature sensitivity is involved in the power measurements
of Section~\ref{power}, we used the same fiber at a constant power of 30.4 mW, and varied the
temperature.
\begin{figure}[h!]
\centering
\includegraphics[width=\linewidth]{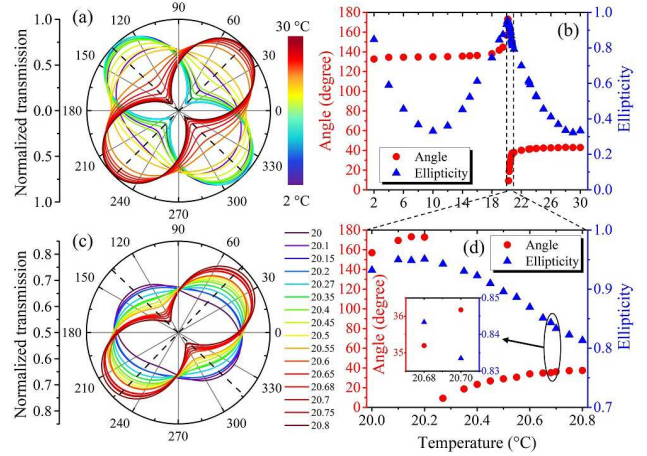}
\caption{\small Projection patterns of the transmitted circularly polarized laser light (at 30.4 mW) through a 17.5 cm
 PM fiber whose 6 cm is exposed to different temperatures
 from a) 2 to 30 $^\circ$C and c) 20 to 20.8 $^\circ$C. 
 b) and d) Ellipticities and angles of the polarization ellipses associated to (a) and (c), respectively.}
 \label{polVStemp}
\end{figure}
A short piece (6 cm) of {Corning} PM fiber in Fig.~\ref{setup} is exposed to different
temperatures by attaching it to a temperature-controlled plate {(Peltier cooler)}.
As shown in Fig.~\ref{polVStemp}(a), the polarization state of the transmitted light gradually changes by increasing the temperature from 2 to 30 $^\circ$C.
  The calculated ellipticities and angles of the polarization ellipses associated to Fig.~\ref{polVStemp}(a) are depicted in Fig.~\ref{polVStemp}(b) as a function of temperature.
The discontinuity in the angle of ellipse at about 20 $^\circ$C in Fig.~\ref{polVStemp}(b) arises from the transition of polarization from circularly (at 20 $^\circ$C) to rotated elliptically (at 21 $^\circ$C) polarized light.
The fitting algorithm only defines angles from 0 to 180$^\circ$ to
 the slow axis of the PM fiber and does not unwrap the curve so that a discontinuity appears when the ellipse rotates past 180 degrees.
  Used as a temperature dependent sensor, placing the system at this transitional location by tuning the fiber length would allow the most sensitivity.
 The sensitivity at the turning point is experimentally measured  by increasing the temperature from 20 $^\circ$C to 20.8 $^\circ$C in
  very small increments as shown in Fig.~\ref{polVStemp}(c).
 {As can be seen, a significant transition of the polarization state of
   light is occurred for only 0.8 $^\circ$C change in temperature of the fiber.}
{The ellipticities and the angles of polarization ellipses at different temperatures from 20 to 20.8 $^\circ$C are depicted in
Fig.~\ref{polVStemp}(d).}
The inset shows that the difference between temperatures of 20.68 and 20.70 $^\circ$C is clearly resolved.\\
It is not necessary to measure the full polarization ellipse to determine the temperature.
 {One can simply measure the transmission of light through a polarizer (Fig.~\ref{setup}) at a fixed orientation.}
There is an optimum polarizer angle for a given temperature range.
  In this experiment, the optimum angles to get the maximum change in transmission at different temperatures are 46 and
   136 degrees as illustrated in Fig.~\ref{polVStemp}(a) and (c) by the dashed lines.
  The direct measurement of the transmitted light through the polarizer at 46 degrees for different temperatures
 is plotted in Fig.~\ref{trans-temp}.
 {The response \textit{r} to temperature change of $\Delta$T is a relative transmitted power change of
 $\Delta P_t/P_t$, proportional to the fiber length of L: }
  \begin{equation}
  \label{equation}
  r=\dfrac{\Delta P_t}{P_t}\dfrac{1}{\Delta TL}.
\end{equation}
{Considering the two data points at 20.68 and 20.70 $^\circ$C shown in the inset of the Fig.~\ref{trans-temp} and for L=0.6  dm,
 the calculated response \textit{r} to the temperature change is 1.2 $^\circ$C$^{-1}$dm$^{-1}$
  It means that for a given resolution of $\Delta P_t/P_t$ of 1$\%$, we can resolve a temperature change of $\Delta$T=0.008 $^\circ$C for 10 cm of fiber. }

There is a compromise to make between sensitivity and dynamic range.  The longer the fiber, the higher the sensitivity,
and the shorter the dynamic range.  {For the 17.5 cm PM fiber length,}
by increasing the temperature from 2 to 30 $^\circ$C, the transmission changes periodically (Fig.~\ref{trans-temp})~\cite{Eickhoff81}.
 This makes the sensor
 impractical for temperatures below 10 $^\circ$C in this case.
 In Fig.~\ref{trans-temp}, the transmission of light through the PM fiber at temperatures below 10 $^\circ$C 
 (highlighted pink region) is very similar to that of temperatures above 10 $^\circ$C (highlighted green region).
 This ambiguity could be avoided by shortening the exposed length of the PM fiber.
   
\subsection{Power-temperature calibration}

\begin{figure}[h!]
\centering
\includegraphics[width=\linewidth]{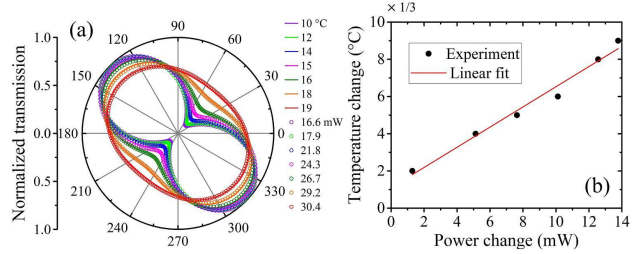}
\caption{\small a) The transmission of
 circularly polarized light at the power of 30.4 mW through the PM fiber at specific temperatures (solid lines) and the
  transmission of circularly polarized light at specific powers through the PM fiber at temperature of 19 $^\circ$C (circles).
b) Changes in temperature of the PM fiber versus changes in power of light passing through the fiber calculated from the legend of (a).}
\label{comparison}
\end{figure}
The strong correlation between the polarization of light at different powers of light (Section~\ref{power}) and different temperatures of the PM fiber
(Section~\ref{interpretation}) is illustrated in Fig.~\ref{comparison}(a).
{It explains the fiber core heating
 through the minuscule absorption of the light in the fused silica fiber.
 This correlation is plotted as a calibration curve in Figure~\ref{comparison}(b).
 As the exposed length of the fiber is different in
  Sections~\ref{power} and \ref{interpretation} 
  (the whole 17.5 cm PM fiber is heated by the power of light
   passing through it while 6 cm of fiber is attached to the temperature-controlled plate),
   we have to normalize the results to the exposed lengths.
   Considering the exposed length of the PM fiber in Section~\ref{interpretation} is almost 1/3 of that in
    Section~\ref{power},
    the temperature changes in Fig.~\ref{comparison}(b) is divided by 3 to keep the response \textit{r} and relative transmission of $\Delta P_t/P_t$ in Eq.~\ref{equation} unchanged.  }

As shown by black dots in Fig.~\ref{comparison}(b), 13 mW change in the power of light passing through the
 PM fiber corresponds to about 2.3 $^\circ$C of change in temperature of the fiber. In other words, a slope of 0.18  $^\circ$C/mW given
 by the linear fit in Fig.~\ref{comparison}(b) {shows that 1 mW change in the power of a CW laser light passing through a 17.5 cm of PM fiber heats the fiber core by 0.18 $^\circ$C}.

\subsection{Response time of power/temperature changes}
\begin{figure}[h!]
\centering
\includegraphics[width=0.4\textwidth]
{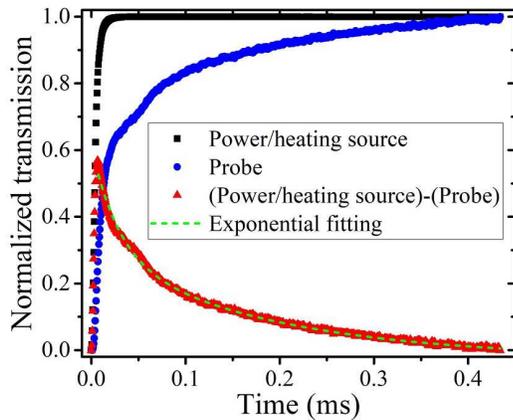}
\caption{\small The response curve of the sensor as measured by 1550 nm laser (probe) to a step function change of the 980 nm laser (power/heating source). }
\label{responseTime}
\end{figure}

The response time of the fiber core temperature to changes in optical power passing through the fiber is
measured by analyzing the step function response of the polarization change.
The power meter in Fig.~\ref{setup} is replaced by a fast detector connected to an oscilloscope.
The lights from two laser diodes at 980 nm (power/heating source) and 1550 nm (probe)
 are combined through a wavelength division multiplexer (WDM) and sent to the QWP in Fig.\ref{setup}.
 The probe power is fixed while the current of the power/heating source is modulated to create square pulses as shown in Fig.~\ref{responseTime}
 by black square symbols as a step function from 0 to 40 mW which is normalized to 1.
 The response curve of the sensor to this step function is measured by recording the transmission of the 1550 nm laser light through a polarizer
 at an optimized angle, and plotted as blue circle symbols.
 The red curve in Fig.~\ref{responseTime} is the  plot of the difference between the power/heating source and the probe,
 used to calculate the actual response time.
Best {fit to this curve} is with two exponentials of 1/e values of  0.028 and 0.187 ms corresponding to cut-off frequencies of 35.7 and 5.3 kHz, respectively.
The faster time constant corresponds to the conduction from core to cladding.
The slower time constant corresponds to the conduction from the fiber to the surroundings.

{A very simple and sensitive fiber optical length stabilization can be
devised based on the birefringence properties of PM fibers and the fast thermal response of the fiber core.
An example of application is where the two arms of a Michelson type interferometer
are made of PM fiber, and have to be stabilized.
It can be done by monitoring the fiber temperature
through polarization modification of an initially circularly polarized beam
launched through the fiber,
and correcting a change in temperature by adjusting the power of a linearly polarized beam of another wavelength sent through the fiber.}

\section{conclusion}
{The birefringence properties of polarization maintaining fibers and the fast thermal response of the fiber core lend themselves to numerous applications.
Beyond the obvious ones such as temperature and power monitoring, saturable absorption and
fiber interferometer stabilization are other possible applications.}

%\bibliography{c:/bib/ad,c:/bib/en,c:/bib/oz,c:/bib/ref-Hanieh}

\end{document}